\documentclass[aps,prb,twocolumn,superscriptaddress,showpacs,10pt,floatfix]{revtex4-2}

\usepackage{natbib}
\usepackage{bibentry}
\usepackage{amsfonts}
\usepackage{amsmath}
\usepackage{braket}
\usepackage{mathtools}
\usepackage{rotating}
\usepackage{color}
\usepackage{framed}
\usepackage{enumerate}
\usepackage{graphicx}
\usepackage{float}
\usepackage{hyperref}
\hypersetup{
    colorlinks=true,
    linkcolor=blue,    
    urlcolor=blue,
    citecolor=blue
    }
\usepackage{tikz}
\usepackage{circuitikz}
\usepackage{tabularx}
\usepackage{multirow}
\usepackage{csquotes}

\newcommand{\vM}{\mathbf{M}}
\newcommand{\ve}{\mathbf{e}}

\definecolor{blue-1}{rgb}{0.756,0.812,0.852}

\begin{document}
\title{Finite-frequency admittance and noise of a helical edge coupled to a magnet}
\author{Oliver Franke}
\affiliation{Dahlem Center for Complex Quantum Systems, Halle-Berlin-Regensburg Cluster of Excellence CCE, and Physics Department, Freie Universit\"at Berlin, Arnimallee 14, 14195 Berlin, Germany}

\author{Paula Koll}
\affiliation{Dahlem Center for Complex Quantum Systems, Halle-Berlin-Regensburg Cluster of Excellence CCE, and Physics Department, Freie Universit\"at Berlin, Arnimallee 14, 14195 Berlin, Germany}

\author{Peter G.~Silvestrov}
\affiliation{Institut f\"ur Mathematische Physik, Technische Universit\"at Braunschweig, D-38106 Braunschweig, Germany}

\author{Piet W. Brouwer}
\affiliation{Dahlem Center for Complex Quantum Systems, Halle-Berlin-Regensburg Cluster of Excellence CCE, and Physics Department, Freie Universit\"at Berlin, Arnimallee 14, 14195 Berlin, Germany}

\begin{abstract}
The exchange coupling of the helical edge state of a quantum spin-Hall insulator with an easy-plane magnet has no effect on its DC electrical conductance if the magnet's anisotropy axis is aligned with the spin quantization axis of the helical edge state [\citeauthor{Meng2014-rs}, Phys. Rev. B 90, 205403 (2014)]. We here calculate the AC conductance $G_V(\omega)$ and the noise power $S_V(\omega)$ in the presence of a DC bias $V$. While both take the universal values $G_V({\omega = 0}) = e^2/h$ and $S_V(\omega = 0) = 4 e^2 k_{\rm B} T/h$ in the zero-frequency limit, $G_V(\omega)$ and $S_V(\omega)$ are quickly suppressed for finite $\omega$, so that low-frequency transport is effectively noiseless.
\end{abstract}

\maketitle

\newpage

\section{Introduction}
\label{sec:introduction}

In an easy-plane magnet, the magnetization aligns in a plane and its energy is, to very good approximation, independent of direction.
The order parameter of an easy-plane magnet thus has a $U(1)$ degree of freedom, which is mathematically analogous to the order parameter of a superconductor \cite{Halperin1969-lc,Sonin1978-xf}. Based on this analogy \cite{Konig2001-al,Nogueira2004-uw}, easy-plane-magnet-based analogs of various forms of ``superfluid'' ({\em i.e.}, dissipationless) spin transport were proposed \cite{Takei2014-lm,Takei2015-dt,Flebus2016-nk,Evers2020-ci}.
Similar effects have been predicted for spin transport through antiferromagnetic insulators \cite{Takei2014-pq}.
The coupling of magnetization dynamics to spin currents in normal metals via ``spin pumping'' (the generation of spin currents by a precessing magnetization \cite{Tserkovnyak2002-ax}) and ``spin torque'' (the exertion of a torque by an electronic spin current  \cite{Berger1996-xi,Slonczewski1996-is}) allows for a rich dynamics in magneto-electronic circuits \cite{Silva2008-hm}. 

\begin{figure}
    \centering
    \includegraphics[width=1\linewidth]{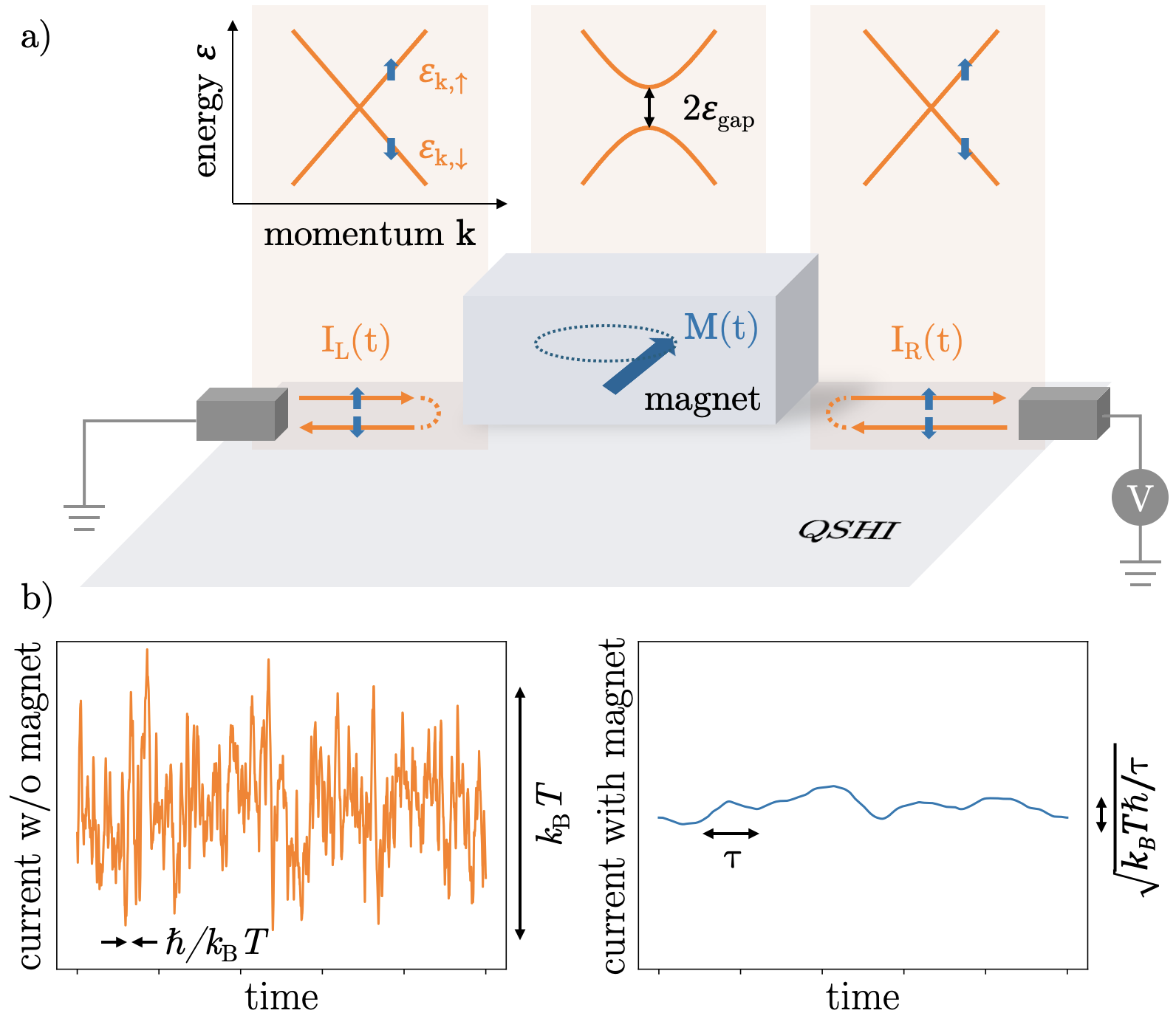}
    \caption{Schematic illustration of a magnet coupled to the helical edge states of a quantum spin Hall insulator (panel a). The exchange coupling to the magnet opens a gap in the spectrum of the helical edge, causing backscattering of electrons accompanied by a spin-flip. This scattering transfers angular momentum to the magnet, driving it out of the easy plane and, hence, inducing a precession of its magnetization, which, in turn, pumps a spin current into the helical edge. Thermal fluctuations in the current carried by incoming electrons are transferred to the magnet and cause small fluctuations in the out-of-plane canting angle and, hence, in the precession frequency $\Omega$. These small fluctuations cause a pumped current temporally correlated with the fluctuations of the incident electron current. Due to the magnet's finite response time $\tau$, high-frequency components of the noise are suppressed. This frequency filtering is illustrated in the two current traces in the bottom panel b). The bottom left panel shows a current trace with correlations $\langle I(\omega)^2\rangle \propto \hbar \omega (e^{\hbar \omega/k_{\rm B}T}-1)^{-1}$, corresponding to Eq.\ (\ref{eq:fdt}) with $G_0(\omega) = e^2/h$ after removing the zero-point fluctuations, whereas the bottom right panel includes the additional low-pass factor $(1-i\omega\tau)^{-1}$ in the presence of a magnet, see Eq.\ (\ref{eq:Ginfinity}).}
    \label{fig:geometry}
\end{figure}

The combination of an easy-plane magnetic insulator with the helical edge of a two-dimensional quantum spin-Hall insulator \cite{Kane2005-pl, Kane2005-vw, Bernevig2006-uk, Bernevig2006-xc, Konig2007-dc} makes the analogy between easy-plane magnetism and superconductivity even more striking and gives additional robustness to the coupling between electronic and magnetic degrees of freedom \cite{Meng2014-rs}. The simplest such geometry is shown schematically in Fig.~\ref{fig:geometry}. It consists of a helical edge exchange-coupled to a magnet with a spatially uniform magnetic moment. The coupling to the magnet opens a gap in the spectrum of the helical edge, causing electrons to backscatter, whereby an angular momentum $\hbar$ is transferred for every electron reflected from the magnet. In turn, helical edge states have a perfect spin-momentum locking, so that a spin current pumped by a precessing magnetization implies a charge current along the helical edge \cite{Qi2008-yf}. For such a setup it was pointed out by \citet*{Meng2014-rs} that an applied voltage $V$ along the helical edge leads to a precession of the magnetization with a precession frequency given by the ``Josephson relation'' $\hbar \Omega = e V$ if the spin quantization axes of the helical edge and the easy-plane magnet are aligned. The precessing magnetic moment pumps a charge current through the helical edge, which has precisely the same magnitude as if there were no magnet-induced backscattering in the helical edge \cite{Meng2014-rs,Silvestrov2016-sp}. As a result, the DC conductance $G_V(\omega = 0) = e^2/h$ of the helical edge is the same, both with and without coupling to the easy-plane magnet.

In a previous article, Recher and two of the authors argued that there is a difference between a helical edge with and without a coupling to an easy-plane magnet if one considers the current fluctuations instead of the average current \cite{Silvestrov2016-sp}: If the magnet-induced gap is at the Fermi energy of the helical edge, the current pumped by the precessing magnet $I_{\rm pump}$ is noiseless, whereas the current in a helical edge without coupling to a magnet has the characteristic thermal noise power $S_{\rm ballistic} = (4e^2/h) k_{\rm B} T$ of a one-dimensional ballistic conductor. Reference \onlinecite{Silvestrov2016-sp} explains the absence of noise using a picture, in which the current was carried effectively by charge carriers far below the Fermi level, which have an occupation that is not subject to fluctuations. In a general setting, the absence of noise reported in Ref.~\onlinecite{Silvestrov2016-sp} follows from the observation that the current is proportional to the precession frequency, $I = e \Omega/2 \pi$, and that the precession frequency $\Omega$ is not subject to fluctuations in the limit of a macroscopic magnet. On the other hand, the absence of thermal noise is in violation of the Nyquist theorem, according to which zero-frequency noise power and the DC linear-response conductance $G_0$ are related as \cite{Johnson1928-mo, Nyquist1928-fk}
\begin{equation}
    S_0(\omega = 0) = 4 k_{\rm B} T G_0(\omega = 0),
  \label{eq:jnnoise}
\end{equation}
where the subscript ``0'' denotes the equilibrium noise in the absence of an applied voltage.

In this article, we resolve this apparent contradiction by considering the conductance $G$ and the noise power $S$ as a function of frequency $\omega$. We show that the precession frequency $\Omega$ of the magnet exhibits slow fluctuations on the time scale \cite{Meng2014-rs}
\begin{align}
  \frac{1}{\tau} = 
  \frac{\hbar \gamma \mathcal{R}}{2 \pi M_{\rm s}}
  \left( \frac{\mathrm{d} \Omega}{\mathrm{d} \theta} \right)_{\!\! \theta = 0},
\label{eq:tau}
\end{align}
where $\gamma$ is the gyromagnetic ratio, $M_{\rm s}$ the total magnetic moment, $\mathcal{R} \in [0, 1]$ a reflection coefficient, and $d\Omega/d\theta$ the derivative of the precession frequency to the out-of-plane canting angle $\theta$, taken at the equilibrium value $\theta = 0$. 
Since the pumped current is proportional to $\Omega$, these fluctuations of the precession frequency restore full thermal fluctuations $S_0(\omega=0) = S_{\rm ballistic}$ of the pumped current in the zero frequency limit. However, as the time scale $\tau$ is extensive \cite{Meng2014-rs} --- it is proportional to the size of the magnet through $M_{\rm s}$ ---, the zero frequency limit is reached for exceedingly small frequencies $\omega$ only. For $\omega \tau \gg 1$ the pumped current is noiseless, consistent with Ref.~\onlinecite{Silvestrov2016-sp}. Since the noise power in the absence of a magnet $S_{{\rm ballistic}}$ is approximately frequency independent for frequencies $\omega \lesssim k_{\rm B} T/\hbar$, current noise indeed distinguishes between a helical edge with and without coupling to an easy-plane magnet, albeit not in the limit of  strictly zero frequency. This is illustrated in the bottom panels of Fig.~\ref{fig:geometry}, which give typical current traces of a ballistic conductor and a helical edge coupled to a magnet.

By explicit calculation of the finite-frequency conductance $G_0(\omega)$ and the finite-frequency equilibrium noise power $S_0(\omega)$, we show that the fluctuation-dissipation theorem \cite{Blanter2000-tn}
\begin{equation}
  \label{eq:fdt}
  S_0(\omega) = 2 \hbar \omega \coth(\hbar \omega/2 k_{\rm B} T)\, \mbox{Re}\, G_0(\omega) 
\end{equation}
is indeed satisfied for all frequencies smaller than the Thouless frequency, $|\omega| \ll \omega_{\rm Th}$, for this system. 
Further, we also consider the case of an applied DC voltage bias and find that admittance $G_V(\omega)$ and noise power $S_V(\omega)$ at bias voltage $V$ are independent of $V$ except for a change in the response time $\tau$ corresponding to the finite canting angle $\theta$ at bias $V$ --- consistent with the absence of shot noise reported in Ref.\ \onlinecite{Silvestrov2016-sp}. 

The remainder of this article is structured as follows: In Sec.\ \ref{sec:magnethelicaledge} we first describe spin currents carried by helical edge states and the magnet that couples to them in a scattering framework. This Section considers the case that there is no DC bias, so that the precession frequency $\Omega$ is zero on the average. With the help of the scattering theory of Sec.\ \ref{sec:magnethelicaledge}, we then calculate the linear response to an AC voltage in Sec.~\ref{sec:admittance} and the noise power $S_0(\omega)$ in Sec.~\ref{sec:noise}. In Sec.\ \ref{sec:dc} we  show that the admittance $G_V(\omega)$ and current noise $S_V(\omega)$ in the presence of a DC bias $V$ can be obtained from the admittance $G_0(\omega)$ and noise $S_0(\omega)$ without DC bias by a simple transformation to a reference frame for the electron spin that precesses at frequency $\Omega = eV/\hbar$. We conclude in Sec.~\ref{sec:discussion}. While the derivations in the main text rely on the $U(1)$ symmetry of the magnet only, we provide an explicit expression for the anisotropy energy in App.\ \ref{sec:anisotropy} as an instructive example and to give a numerical estimate for the response time $\tau$. In App.\ \ref{sec:heatbath} we calculate the conductance and the noise power for the case that the magnet is subject to additional damping, whereas App.\ \ref{sec:dcbias} provides additional details regarding the admittance and noise power in the presence of a DC bias $V$.

\section{Easy-plane magnet and helical edge}
\label{sec:magnethelicaledge}
We consider electrons in a helical edge state that are exchange-coupled to a magnetic moment $\vM(t)$ with easy-plane anisotropy, see Fig.~\ref{fig:geometry}. The magnitude $|\vM(t)| = M_{\rm s}$, $M_{\rm s}$ being the saturated moment, and we refer to App.~\ref{sec:anisotropy} for details regarding units and notation. The hard axis $\ve_z$ of the magnet is aligned with the spin quantization axis of the helical edge state. If $\vM$ is canted out of the $xy$ plane with canting angle $\theta$, it precesses with a precession frequency $\Omega(\theta)$. For an easy-plane magnet, one has $\Omega(0) = 0$ with a finite derivative $d\Omega/d\theta$ at $\theta = 0$. A quantum magnet has discrete precession frequencies $\Omega$ spaced by $\Delta \Omega = \hbar \gamma/M_{\rm s} (d\Omega/d\theta) = 2 \pi/{\cal R} \tau$ for out-of-plane canting angles $\theta \ll 1$.Since we treat the precession frequency $\Omega$ as an effectively continuous variable, we require that $\hbar \Delta \Omega \ll k_{\rm B} T$, or equivalently, for ${\cal R} \sim 1$, 
\begin{equation}
  \frac{\hbar}{\tau} \ll k_{\rm B} T.
  \label{eq:cond1}
\end{equation}

In the helical edge, electrons with spin up (measured along the $z$ direction) move to the right, whereas electrons with spin down move to the left. Below, we use the short-hand notation $\hat a_{\rm L}(\varepsilon)$ and $\hat a_{\rm R}(\varepsilon)$ for annihilation operators of electrons with spin up or down incident on the region coupled to the magnet from the left or right, respectively. Similarly, we use $\hat b^{\Omega}_{\rm L}(\varepsilon)$ and $\hat b_{\rm R}^{\Omega}(\varepsilon)$ for annihilation operators of electrons with spin up, down moving away from the region coupled to the magnetic moment precessing at frequency $\Omega$ on the left or right side. In thermal equilibrium, the electrons incident on the scattering region follow the Fermi-Dirac distribution               
\begin{equation}
  \label{eq:aa}
  \langle \hat{a}_{\alpha}^\dag(\varepsilon) \hat{a}_{\beta}(\varepsilon') \rangle = \delta_{\alpha \beta} \delta(\varepsilon - \varepsilon') f_{\alpha}(\varepsilon), \quad \alpha, \beta = \mathrm{L, R},
\end{equation}
with
\begin{equation}
  f_{\alpha} = \frac{1}{e^{(\varepsilon - \mu_{\alpha})/k_{\rm B} T} + 1}
\end{equation}
and $\mu_{\rm L}$ and $\mu_{\rm R}$ the chemical potentials of the left and right reservoirs, respectively.

The electron operators for the outgoing scattering states can be expressed in terms of those of the incoming scattering states via the scattering amplitudes $r(\varepsilon)$, $t(\varepsilon)$, $t'(\varepsilon)$, and $r'(\varepsilon)$. For a magnetization that precesses at a constant frequency $\Omega$, the relation between the operators for outgoing and incoming scattering states reads \cite{Silvestrov2016-sp} 
\begin{align}
  \label{eq:abrt}
\begin{split}
  \hat b_{\rm L}^{\Omega}(\varepsilon- \tfrac12 \hbar \Omega) &=
  r(\varepsilon) \hat a_{\rm L}(\varepsilon + \tfrac12 \hbar \Omega)
  + t'(\varepsilon) \hat a_{\rm R}(\varepsilon-\tfrac12 \hbar \Omega), \\
  \hat b_{\rm R}^{\Omega}(\varepsilon+\tfrac12 \hbar \Omega) &=
  t(\varepsilon) \hat a_{\rm L}(\varepsilon+\tfrac12 \hbar \Omega)
  + r'(\varepsilon) \hat a_{\rm R}(\varepsilon - \tfrac12 \hbar \Omega).
\end{split}
\end{align}
The energy shift by $\hbar \Omega$ upon backscattering can be understood by moving to a frame that co-rotates with the precessing magnetization. In the co-rotating frame, scattering from the magnet does not change the energy of the carriers. Shifting back to the original frame gives an energy shift $\pm \hbar \Omega/2$ for spin up and down, respectively.

In this Section and in Secs.\ \ref{sec:admittance} and \ref{sec:noise} we consider a precession frequency $\Omega$ that is zero on the average but fluctuates in time. The fluctuations of $\Omega$ take place on the time scale $\min(1/\omega,\tau)$ if an AC bias at frequency $\omega$ is applied and at time scale $\tau$ for equilibrium fluctuations. We assume that the time scale for fluctuations of $\Omega$ is long compared to the typical dwell time $\tau_{\rm dwell}$ of electrons in the region of the helical edge that is coupled to the magnet, so that Eq.~\eqref{eq:abrt} holds locally in time. This amounts to the conditions
\begin{equation}
  \omega,\ 1/\tau \ll \omega_{\rm Th},
  \label{eq:cond2}
\end{equation}
where $\omega_{\rm Th} \sim 1/\tau_{\rm dwell}$ is the Thouless frequency. Further, we also require that the root-mean-square magnitude of the fluctuations of the precession frequency, ${\rm rms}\, \Omega$, and that the frequency $\omega$ at which current fluctuations are considered be small in comparison to $\omega_{\rm Th}$,
\begin{equation}
  \omega,\ {\rm rms}\, \Omega \ll \omega_{\rm Th}.
  \label{eq:cond3}
\end{equation}
Since $\omega_{\rm Th}$ sets the scale for the energy dependence of the reflection and transmission amplitudes, the condition (\ref{eq:cond3}) implies that we may neglect shifts of the energy arguments of the reflection and transmission amplitudes by an amount $\sim \hbar \Omega$ or $\sim \hbar \omega$.
If electrons are partially transmitted and motion in the part of the helical edge coupled to the magnet is ballistic, one has $\omega_{\rm Th} \sim v/L$, where $v$ is the velocity of electrons in the helical edge and $L$ is the linear size of the magnet along the edge. If all electrons are reflected from the magnet in the entire energy window of interest, one has $\omega_{\rm Th} \sim \varepsilon_{\rm gap}/\hbar$, where $\varepsilon_{\rm gap}$ is the magnet-induced excitation gap in the helical edge.

In Sec.\ \ref{sec:noise} we will show that ${\rm rms}\, \Omega \sim (k_{\rm B} T/\hbar \tau)^{1/2}$ for thermal fluctuations, so that the conditions (\ref{eq:cond1}) and (\ref{eq:cond3}) imply the inequality (\ref{eq:cond2}). Since the relaxation time $\tau$ scales proportional to the volume of the magnet, see Eq.\ (\ref{eq:tau}), the estimate ${\rm rms}\, \Omega \sim (k_{\rm B} T/\hbar \tau)^{1/2}$ also implies that the condition (\ref{eq:cond3}) is typically (very generously) fulfilled for a macroscopic magnet.

The current $\hat I_{\rm L}(t)$ in the left terminal reads
\begin{align}
\begin{split}
  \hat I_{\rm L}(t) =&\ 
   \frac{e}{h} \int \mathrm{d}\varepsilon \int \mathrm{d}\varepsilon'
  e^{i(\varepsilon - \varepsilon') t / \hbar} \\ 
  &\, \times
  [\hat a_{\rm L}^{\dagger}(\varepsilon) \hat a_{\rm L}(\varepsilon')
  - \hat b_{\rm L}^{\Omega\dagger}(\varepsilon) 
  \hat b_{\rm L}^{\Omega}(\varepsilon')].
\end{split}
\end{align}
Expanding this expression to linear order in the precession frequency $\Omega$ gives an expression for $\hat I_{\rm L}(t)$ in terms of operators $\hat b_{\rm L,R}(\varepsilon) \equiv \hat b_{\rm L,R}^{\Omega = 0}(\varepsilon)$ for scattering states at a fixed ({\em i.e.}, non-precessing) magnetization,
\begin{align}
\begin{split}
  \hat I_{\rm L}(t) =&\
   \frac{e}{h} \int \mathrm{d}\varepsilon \int \mathrm{d}\varepsilon'
  e^{i(\varepsilon - \varepsilon') t / \hbar} \\ 
  &\, \times
  [\hat a_{\rm L}^{\dagger}(\varepsilon) \hat a_{\rm L}(\varepsilon')
  - \hat b_{\rm L}^{\dagger}(\varepsilon) 
  \hat b_{\rm L}(\varepsilon')] \\
  &\, + \hat I_{\rm pump}(t),
\end{split}
\label{eq:IL}
\end{align}
where
\begin{align}
  \label{eq:Ipump}
  \hat I_{\rm pump}(t) =&\,
  \frac{e}{2 \pi} \Omega \mathcal{R}
\end{align}
is the current ``pumped'' by the precessing magnetic moment. Here,
\begin{align}
  \mathcal{R} = 
  \int \mathrm{d}\varepsilon
  \left( - \frac{\partial f}{\partial \varepsilon} \right)
  |r(\varepsilon)|^2
\label{eq:r}
\end{align}
is the reflection coefficient of the magnet.
Equation (\ref{eq:Ipump}) simplifies to the relation $\hat I_{\rm pump} = e \Omega/2 \pi$ for the case $\mathcal{R} = 1$, which describes a helical edge coupled to a macroscopic magnet, with the Fermi energy inside the magnet-induced gap.
For the low frequencies $|\omega| \ll \omega_{\rm Th}$ we are interested in, the current in the helical edge is conserved. We therefore write
\begin{equation}
  \label{eq:currentconservation}
  \hat I(t) \equiv
  \hat I_{\rm L}(t) = \hat I_{\rm R}(t).
\end{equation}

Because $\hat I_{\rm pump}$ is manifestly linear in $\Omega$, in Eq.~(\ref{eq:Ipump}) we took the equilibrium expectation value of the products of creation and annihilation operators for electrons in the scattering states. This is consistent with the fact that a current pumped by a magnetization that precesses at a constant rate $\Omega$ is not subject to thermal noise. (That the pumped current is noiseless for a constant precession frequency $\Omega$ is most easily seen by shifting to a reference frame co-rotating with the precessing magnetization. The shift to such a reference frame comes with a uniform offset to the current in the helical edge and a change of the chemical potentials in the two reservoirs, neither of which affects the thermal noise. For more details, see App.\ \ref{sec:dcbias}.) Fluctuations of $\hat I_{\rm pump}$ and, hence, of $\hat I$ arise from fluctuations of the precession frequency $\Omega$. Inclusion of fluctuations of $\Omega$ differentiates the present calculation from that of Ref.~\onlinecite{Silvestrov2016-sp}, where it was assumed that the precession frequency $\Omega$ of a macroscopic magnet is strictly constant.

The time-dependence of the out-of-plane component $M_z$ of the magnetization, including its fluctuations inherited from the scattered electrons, follows from the continuity equation for angular momentum,
\begin{align}
\begin{split}
  \dot{M_z}(t) =&\
  \frac{\gamma}{4 \pi}
  \int \mathrm{d}\varepsilon \int \mathrm{d}\varepsilon' 
  e^{i (\varepsilon - \varepsilon')t/\hbar} \\ &\, \times 
  [\hat a^{\dagger}_{\rm L}(\varepsilon) \hat a_{\rm L}(\varepsilon')
  + \hat b^{\Omega\dagger}_{\rm L}(\varepsilon) \hat b_{\rm L}^{\Omega}(\varepsilon') \\
  &\, \quad \ - \hat a^{\dagger}_{\rm R}(\varepsilon) \hat a_{\rm R}(\varepsilon')
  - \hat b^{\Omega\dagger}_{\rm R}(\varepsilon) \hat b_{\rm R}^{\Omega}(\varepsilon')].
\end{split}
\label{eq:dotMz}
\end{align}
Again expanding to first order in $\Omega$ and taking expectation values for terms that are manifestly linear in $\Omega$, we find
\begin{align}
\begin{split}
  \dot{M_z}(t) =&\
  \frac{\gamma}{4 \pi}
  \int \mathrm{d}\varepsilon \int \mathrm{d}\varepsilon' 
  e^{i (\varepsilon - \varepsilon')t/\hbar} \\ &\, \times 
  [\hat a^{\dagger}_{\rm L}(\varepsilon) \hat a_{\rm L}(\varepsilon')
  + \hat b^{\dagger}_{\rm L}(\varepsilon) \hat b_{\rm L}(\varepsilon') \\ 
  &\, \quad \ - \hat a^{\dagger}_{\rm R}(\varepsilon) \hat a_{\rm R}(\varepsilon')
  - \hat b^{\dagger}_{\rm R}(\varepsilon) \hat b_{\rm R}(\varepsilon')] \\
  &\, - \frac{\hbar \gamma}{e} \hat I_{\rm pump}(t).
\end{split}
\label{eq:dotMzz}
\end{align}
Since $M_z = M_{\rm s} \sin \theta$, for small $\Omega$ and canting angle $\theta$, one has
\begin{equation}
  \label{eq:MzOmega}
  \Omega = \frac{M_{z}}{M_{\rm s}} \left. \left( \frac{\mathrm{d} \Omega}{\mathrm{d} \theta} \right)\right|_{\theta \to 0}.
\end{equation}

Equations (\ref{eq:Ipump}), (\ref{eq:dotMzz}), and (\ref{eq:MzOmega}) form a coupled set, from which the precession frequency $\Omega$ and the pumped current $\hat I_{\rm pump}$ can be obtained. A solution of these equations is most conveniently obtained by Fourier transform to the time argument $t$, which gives
\begin{align}
\label{eq:IpumpResult}
  \hat I_{\rm pump}(\omega)=&\ \frac{e}{2(1 - i \omega \tau)} 
  \int \mathrm{d}\varepsilon \\ &\,\times
  [ \hat a^{\dagger}_{\rm L}(\varepsilon) \hat a_{\rm L}(\varepsilon+\hbar \omega)
  + \hat b^{\dagger}_{\rm L}(\varepsilon) \hat b_{\rm L}(\varepsilon+\hbar \omega) \nonumber \\ &\, \quad \
  - \hat a^{\dagger}_{\rm R}(\varepsilon) \hat a_{\rm R}(\varepsilon+\hbar \omega)
  - \hat b^{\dagger}_{\rm R}(\varepsilon) \hat b_{\rm R}(\varepsilon+\hbar \omega) ], \nonumber
\end{align}
where $\tau$ is the characteristic response rate of the out-of-plane magnetization amplitude to the current carried by the helical edge as given in Eq.~(\ref{eq:tau}). Since the magnetic moment $M_{\rm s}$ is extensive, the response time $\tau$ is an extensive quantity as well. An explicit expression for $\tau$ in terms of microscopic quantities is given for the case of an anisotropy energy quadratic in $M_z$ in App.~\ref{sec:anisotropy}.

Upon substituting Eq.~(\ref{eq:IpumpResult}) into Eq.~(\ref{eq:IL}) and using current conservation (\ref{eq:currentconservation}), we arrive at
\begin{align}
\begin{split}
  \hat I(\omega) 
  =&\, e \int \mathrm{d}\varepsilon \bigg[\hat a^{\dagger}_{\rm L}(\varepsilon) \hat a_{\rm L}(\varepsilon +\hbar \omega) \\
  &\, \quad \quad \quad \ + \frac{i \omega \tau}{1 - i \omega \tau} \hat b^{\dagger}_{\rm L}(\varepsilon) \hat b_{\rm L}(\varepsilon+\hbar \omega) \\
  &\, \quad \quad \quad \ - \frac{1}{1 - i \omega \tau}
  \hat a^{\dagger}_{\rm R}(\varepsilon) \hat a_{\rm R}(\varepsilon+\hbar \omega) \bigg].
\end{split}
\label{eq:Iomega}
\end{align}
In the DC limit $\omega \to 0$, Eq.~(\ref{eq:Iomega}) reproduces the current
\begin{equation}
  \hat I_{\rm ballistic}(\omega = 0) = e \int d\varepsilon \left[\hat a^{\dagger}_{\rm L}(\varepsilon) \hat a_{\rm L}(\varepsilon) - \hat a^{\dagger}_{\rm R}(\varepsilon) \hat a_{\rm R}(\varepsilon) \right]
\end{equation}
on the operator level, implying that  not only the average current $I(\omega = 0)$, but also of the zero-frequency equilibrium current noise $S_0(\omega =0)$ are equal for a helical edge with and without coupling to an easy-plane magnet, as required by the Johnson-Nyquist expression (\ref{eq:jnnoise}). Below, we now evaluate the average current in linear response to an AC bias and the equilibrium current fluctuations at finite frequency $\omega$.

\section{AC conductance}
\label{sec:admittance}

To calculate the AC linear-response conductance $G_0(\omega)$, we apply a (small) time-dependent voltage 
\begin{equation}
 \delta V_{\alpha}(t) = \int \frac{d\omega}{2 \pi} \delta V_{\alpha}(\omega) e^{-i \omega t}
\end{equation}
to the reservoirs coupled to the helical edge to the left ($\alpha = {\rm L}$) or right ($\alpha = {\rm R}$) of the magnet. In this case, the distribution of the incident electrons is, to first order in $\delta V_{\alpha}(\omega)$, 
\begin{align}
  \label{eq:aaomega}
  \langle \hat{a}_{\alpha}^\dag(\varepsilon) \hat{a}_{\beta}(\varepsilon') \rangle =&\
   \delta_{\alpha \beta} \delta(\varepsilon - \varepsilon') f(\varepsilon) 
  \\ \nonumber &\,
  + \frac{e}{h} \delta_{\alpha \beta}
  \frac{f(\varepsilon) - f(\varepsilon')}{\varepsilon' - \varepsilon}
  \delta V_{\beta}[(\varepsilon - \varepsilon')/\hbar].
\end{align}
The current response may be written in terms of the admittance coefficients $G_{\alpha\beta}(\omega)$ as \cite{Buttiker1996-fc}
\begin{align}
\label{eq:ohm}
  I_{\alpha}(\omega) = \sum_{\beta = {\rm L, R}} G_{\alpha\beta}(\omega)
  \delta V_{\beta}(\omega).
\end{align}
For frequencies $|\omega| \ll \omega_{\rm Th}$, current is conserved, so that there is a well-defined AC conductance $G_0(\omega)$,
\begin{align}
  G_0(\omega) \equiv&\, G_{\alpha{\rm L}}(\omega) = -G_{\alpha{\rm R}}(\omega).
\end{align}
From Eqs.\ (\ref{eq:Iomega}) and (\ref{eq:aaomega}), we find that the AC conductance $G_0(\omega)$ is ({\it cf.} Refs.~\onlinecite{Buttiker1993-jy, Pretre1996-du, Pedersen1998-ze})
\begin{align}
\label{eq:G}
  G_0(\omega) =&\
  \frac{e^2}{h} \int \mathrm{d}\varepsilon
  \frac{f(\varepsilon) - f(\varepsilon + \hbar \omega)}{\hbar \omega}
  \frac{1 - i \omega \tau |t(\varepsilon)|^2}{1 - i \omega \tau}
  \nonumber \\ =&\
  \frac{e^2}{h} \left(1 + \mathcal{R} \frac{i \omega \tau}{1 - i \omega \tau} \right),
\end{align}
where, in the second line, we used the inequality $|\omega| \ll \omega_{\rm Th}$ to further simplify the expression. Equation (\ref{eq:G}) is consistent with the exponential relaxation to the steady-state current $I_{\rm ballistic} = e^2 V/h$ reported in Ref.\ \onlinecite{Meng2014-rs}. In Sec.~\ref{sec:dc} and App.~\ref{sec:dcbias} we show that a very similar result for the admittance is found in the presence of a DC bias.

\begin{figure}
    \centering
    \includegraphics[width=0.5\textwidth]{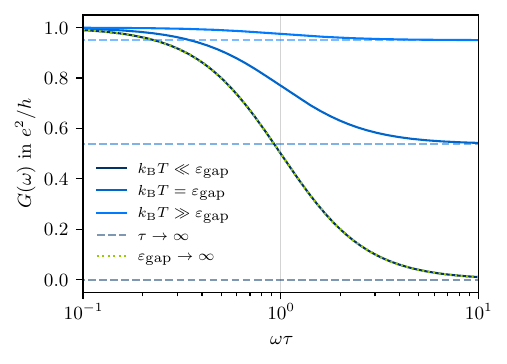}
    \caption{AC conductance $G_0(\omega)$ vs.\ frequency $\omega$ for the case that the magnet-induced gap in the spectrum of the helical edge is of size $2 \varepsilon_{\rm gap}$, with the Fermi energy in the gap center (see text). The solid curves show $G_0(\omega)$ for $k_{\rm B} T/\varepsilon_{\rm gap} = 0.1$, $1$, and $10$ (see legend). The dashed lines indicate the limit $\tau \to \infty$ of Eq.~(\ref{eq:tauinf}). The dotted curve indicates the limit $\varepsilon_{\rm gap}/k_{\rm B} T \to \infty$ of Eq.~(\ref{eq:Ginfinity}). For temperature $k_{\rm B} T \ll \varepsilon_{\rm gap}$, the AC conductance and the thermal noise are strongly suppressed in the frequency range $\tau^{-1} \ll \omega \ll \varepsilon_{\rm gap}/\hbar$.}
    \label{fig:G}
\end{figure}

To illustrate the frequency dependence of the admittance, we consider the case that the exchange coupling of the magnet opens up a gap of size $2 \varepsilon_{\rm gap}$ in the spectrum of the helical edge, with the Fermi energy in the gap center. For the transmission coefficient $|t(\varepsilon)|^2$, this corresponds to $|t(\varepsilon)|^2 = \Theta(|\varepsilon| - \varepsilon_{\text{gap}})$, so that $\mathcal{R} = \tanh(\varepsilon_{\rm gap}/2 k_{\rm B} T)$. We show the corresponding AC conductance $G_0(\omega)$ of Eq.~\eqref{eq:G} in Fig.~\ref{fig:G}. In the limit $\tau \to \infty$ at fixed $\varepsilon_{\rm gap}$, which is appropriate for a macroscopic magnet, since $\varepsilon_{\rm gap}$ is an intrinsic energy scale, whereas $\tau$ is extensive, the AC conductance reduces to
\begin{equation}
  \label{eq:tauinf}
    G_0(\omega) = \frac{e^2}{h} \frac{2}{1 + e^{\varepsilon_{\rm gap}/k_{\rm B} T}},
\end{equation}
independent of frequency $\omega$. The AC conductance is exponentially suppressed $\propto e^{-\varepsilon/k_{\rm B} T}$ in the limit of low temperatures. In the limit $\varepsilon_{\rm gap} \to \infty$, we find
\begin{equation}
     G_0(\omega) = \frac{e^2}{h} \frac{1}{1 - i \omega \tau}.
  \label{eq:Ginfinity}
\end{equation}

\section{Noise power}
\label{sec:noise}

We now compare a calculation of the equilibrium noise power $S_0(\omega)$ with that of the AC conductance $G_0(\omega)$. Hereto, we not only account for the noise from the current carried by incident electrons, but also for the noise from the pumped current and cross terms. (The cross terms are important if both reflection and transmission through the magnet are nonzero.) We assume that the only source of noise driving the magnetization dynamics is the torque exerted by the reflected electrons in the helical edge. We discuss the case that the magnetic moment is additionally coupled to a thermostat in App.~\ref{sec:heatbath}.

The noise power $S(\omega)$ is defined as \cite{Blanter2000-tn}
\begin{align}
\label{eq:noisepower}
    2 \pi \delta(\omega + \omega') S_{\alpha \beta} (\omega) 
    =&\, \langle \hat{I}_{\alpha}(\omega) \hat{I}_{\beta}(\omega') +  \hat{I}_{\beta}(\omega') \hat{I}_{\alpha}(\omega) \rangle 
  \nonumber \\ &\, \mbox{}
- 2 \langle \hat{I}_{\alpha}(\omega) \rangle \langle \hat{I}_{\beta}(\omega') \rangle. 
\end{align}
In the regime where current is conserved, one has
\begin{equation}
  S(\omega) = S_{\alpha\beta}(\omega),
\end{equation}
independent of $\alpha$ and $\beta$. To calculate the equilibrium noise $S_0(\omega)$ at temperature $T$, we insert Eq.~\eqref{eq:Iomega}, calculate the expectation values of the products of creation and annihilation operators using Eq.~(\ref{eq:aa}), and obtain
\begin{align}
\begin{split}
  S_0(\omega) =&\
  \frac{2 e^2}{h} \int \mathrm{d}\varepsilon
  \frac{1 + \omega^2 \tau^2 |t(\varepsilon)|^2}{1 + \omega^2 \tau^2} \\ &\, \times
  [f(\varepsilon) + f(\varepsilon + \hbar \omega)
  - 2 f(\varepsilon) f(\varepsilon + \hbar \omega)].
\end{split}
\label{eq:S}
\end{align}
Using the identity $f(\varepsilon)[1 - f(\varepsilon+\hbar \omega)] + f(\varepsilon + \hbar \omega)[1 - f(\varepsilon)] = [f(\varepsilon) - f(\varepsilon + \hbar \omega)] \coth{(\hbar \omega / 2 k_{\rm B} T)}$ and the inequality $|\omega| \ll \omega_{\rm Th}$, we may further simplify this result as
\begin{equation}
  S_0(\omega) = \frac{e^2 \omega}{\pi} 
  \left(1 - \mathcal{R} \frac{\omega^2 \tau^2}{1 + \omega^2 \tau^2} \right)
  \coth \frac{\hbar \omega}{2 k_{\rm B} T}.
  \label{eq:Sfinal}
\end{equation}
{}From here, one easily verifies that Eqs.~\eqref{eq:G} and \eqref{eq:Sfinal} fulfill the fluctuation-dissipation relation of Eq.~\eqref{eq:fdt}.

We again consider a few limiting cases. For a macroscopic magnet, taking the limit $\tau \to \infty$ at fixed $\omega$, we find that the noise power $S_0(\omega) \to 0$ if there is no direct transmission through the magnet, as was previously found in Ref.~\onlinecite{Silvestrov2016-sp}. For the simple model $|t(\varepsilon)|^2 = \Theta(|\varepsilon| - \varepsilon_{\rm gap})$, corresponding to a magnet-induced spectral gap of size $2 \varepsilon_{\rm gap}$, with the Fermi energy in the gap center, the noise power $\propto e^{-\varepsilon_{\rm gap}/k_{\rm B} T}$ is exponentially suppressed with temperature. However, if we first take the limit $\omega \to 0$ for a finite $\tau$ and use $f(\varepsilon) [1 - f(\varepsilon)] = k_{\rm B} T (-\partial f / \partial \varepsilon)$, Eq.~\eqref{eq:S} reduces to the Johnson-Nyquist noise proportional to $k_{\rm B}T$ in Eq.~\eqref{eq:jnnoise}, irrespective of the transmission probability $|t(\varepsilon)|^2$.

Our calculation is based on a linear expansion in the fluctuations of $\Omega$. We therefore verify that the magnitude of the fluctuations is small compared to the Thouless energy scale. From Eqs.~(\ref{eq:Ipump}) and (\ref{eq:IpumpResult}) we find for the fluctuation size of $\Omega$, {\it i.e.}, the square root of its correlator,
\begin{equation}
  \mbox{rms}\, \Omega = \sqrt{\frac{\gamma k_{\rm B}T }{M_{\rm s}} \left( \frac{\mathrm{d} \Omega}{\mathrm{d} \theta} \right)_{\!\! \theta = 0}}.
\end{equation}
This result is consistent with the average energy of the magnet being equal to $k_{\rm B}T/2$, see App.~\ref{sec:anisotropy}.
The condition $\mbox{rms}\, \Omega \ll \omega_{\rm Th}$ sets a temperature range, for which the expansion in small $\Omega$ is valid. Since $M_{\rm s}$ is proportional to the volume of the magnet, whereas $\omega_{\rm Th} \sim v/L$ scales inversely proportional to its linear size if there is at least partial transmission and ballistic motion in the region coupled to the edge, or $\omega_{\rm Th}$ is $L$-independent if reflection is complete in the energy window of interest, the inequality $\mbox{rms}\, \Omega \ll \omega_{\rm Th}$ holds for sufficiently low temperatures and sufficiently large magnets. This establishes the validity of the small-$\Omega$ expansion of Sec.\ \ref{sec:magnethelicaledge}.

\section{DC bias}
\label{sec:dc}

In the previous Sections, we calculated the admittance $G_0(\omega)$ and the equilibrium noise power $S_0(\omega)$ for a helical edge coupled to an easy-plane magnet in the absence of a DC bias, so that the precession frequency $\Omega$ and the out-of-plane canting angle $\theta$ are zero on the average. We now address the admittance $G_V(\omega)$ and the noise power $S_V(\omega)$ in the presence of an applied DC bias $V$, for which $\Omega$ has small fluctuations around the average value $e V / \hbar$. 

As shown in detail in App.~\ref{sec:dcbias},  by transforming to a reference frame for the spin degree of freedom that rotates at frequency $eV/\hbar$, the DC bias can be eliminated, and the calculation of the admittance and noise power at finite DC bias can be mapped to the calculation at zero DC bias. We thus find that the admittance and the noise are independent of $V$, except for an indirect dependence of the relaxation time $\tau$ on $V$ via the $V$-dependence of the out-of-plane canting angle $\theta$,
\begin{align}
  G_V(\omega) =&\, \left. G_0(\omega) \right|_{\tau \to \tau_V}, \nonumber \\
  S_V(\omega) =&\, \left. S_0(\omega) \right|_{\tau \to \tau_V},
  \label{eq:SV}
\end{align}
with
\begin{equation}
  \frac{1}{\tau_V} = 
  \frac{\hbar \gamma \mathcal{R}}{2 \pi M_{\rm s} \cos \theta_V}
  \left( \frac{\mathrm{d} \Omega}{\mathrm{d} \theta} \right)_{\!\! \theta = \theta_V},
\label{eq:tauV}
\end{equation}
where the finite-bias out-of-plane canting angle $\theta_V$ is the solution of $\hbar \Omega(\theta) = e V$. (Note that Eq.~(\ref{eq:tauV}) simplifies to Eq.~(\ref{eq:tau}) in the limit $V \to 0$, since $\theta_0 = 0$.)
For comparison: For a ballistic one-dimensional channel one also has $S_{V,{\rm ballistic}}(\omega) = S_{0,{\rm ballistic}}(\omega)$ \cite{Blanter2000-tn}. Equation (\ref{eq:SV}) for the noise power of a helical edge coupled to an easy-plane magnet is nevertheless remarkable, since shot noise, the difference of $S_V(\omega)$ and $S_0(\omega)$, is usually associated with the partitioning of charge that arises upon backscattering of charge carriers \cite{Blanter2000-tn}, whereas Eq.~(\ref{eq:SV}) holds irrespective of the reflection coefficient $\mathcal{R}$ of the magnet.

\section{Discussion}
\label{sec:discussion}

In this article, we calculated the noise power $S_V(\omega)$ and the AC conductance $G_V(\omega)$ for a helical edge of a quantum spin-Hall insulator coupled to an easy-plane magnet in the presence of a DC bias $V$. We find that both $G_V(\omega)$ and $S_V(\omega)$ vanish if the frequency $\omega$ exceeds a characteristic relaxation time $\tau$, whereas $G_V(\omega)$ and $S_V(\omega)$ approach universal limits $G_V = e^2/h$ and $S_V = 4 e^2 k_{\rm B} T/h$ if $\omega \ll 1/\tau$. For a macroscopic magnet, one has $\tau \to \infty$, so that $S_V(\omega) \to 0$ for any finite frequency $\omega$, a result that was advertised as ``noiseless'' transport in a previous publication by Recher and two of the authors \cite{Silvestrov2016-sp}. The noise power $S_V(\omega)$ is independent of the DC bias voltage $V$ for all frequencies $\omega$, consistent with the absence of shot noise reported in Ref.\ \onlinecite{Silvestrov2016-sp}. We verified that the equilibrium noise power $S_0(\omega)$ and the AC conductance $G_0(\omega)$ satisfy the fluctuation-dissipation relation --- but this should not come as a surprise.

The $U(1)$ nature of the magnetic order parameter of an easy-plane magnet has stimulated analogies with superconductors \cite{Konig2001-al,Nogueira2004-uw,Takei2014-lm,Takei2015-dt,Flebus2016-nk,Evers2020-ci}. For an easy-plane magnet coupled to a helical edge, this analogy becomes visible in the ``Josephson relation'' $\Omega_V = eV/\hbar$ between the precession frequency of the magnetic moment and the applied bias $V$. As was already pointed out by \citet{Meng2014-rs}, additional relaxation processes for the magnetization dynamics --- phenomenologically described by a Gilbert damping term in the equation of motion for the out-of-plane component $M_z$ of the magnetic moment --- lead to a reduction of the precession frequency and, hence, of the DC conductance, see also App.\ \ref{sec:heatbath}. 

In order to observe the phenomena described here, it is necessary that the characteristic relaxation time $\tau$ of a magnet coupled to a helical edge be smaller than the intrinsic relaxation time $\tau_{\alpha}$ associated with Gilbert damping. Since $\tau$ is extensive, this in practice limits the effects described here to mesoscopic magnets. As an example, for the easy-plane magnet K$_2$CuF$_4$, \citet{Meng2014-rs} estimate \mbox{$\tau^{-1} \approx 1.5 \, \mathrm{nm}^{3} \, \mathrm{ns}^{-1} / V_{\rm M}$}, which should be compared with the estimate $\tau_{\alpha}^{-1} \approx 0.5 \, \mathrm{ns}^{-1}$ (for $\alpha \approx 0.01$). The condition $\tau \ll \tau_{\alpha}$ then implies a maximum volume $V_{\rm M} \sim 3 \, \mathrm{nm}^3$.

In addition to the condition that the magnet is small enough that Gilbert damping plays no role in comparison to the pumped spin current, we assumed that the hard axis of the easy-plane magnet is aligned with the spin quantization axis of the helical edge and that the magnet has no additional easy-axis anisotropy. \citet{Xiao2021-lr} considered classical equations of motion for the magnetic moment of an easy-plane magnet coupled to a helical edge that in principle account for the latter two effects. Understanding the effect of such non-idealities on current fluctuations is an open question. Another important assumption in our calculations is that all frequencies of interest are well below the Thouless frequency $\omega_{\rm Th}$. This is a standard approximation in the field of current-induced magnetization dynamics \cite{Tserkovnyak2005-dt} and we verified that the conditions required for this approximation are satisfied in the present case.

We close with a few remarks on possible implications from our findings. By choosing a magnet of the appropriate size and type, it becomes possible to tune the frequency window, in which the noise is exponentially suppressed, or build a low-pass filter for the charge current. An understanding of the noise and admittance is also relevant to earlier proposals of the system considered here as a microwave resonator or an application in quantum spin-Hall insulator circuits \cite{Meng2014-rs}. As originally suggested in Ref.~\onlinecite{Silvestrov2016-sp}, our results show that not only interferometry \cite{Madsen2021-wx}, but also a measurement of the noise power can be used to distinguish a helical edge coupled to a magnet from one that is not, despite their DC conductances being equal \cite{Madsen2021-wx}.

\section*{Acknowledgments}
We thank Carlo Beenakker, Patrik Recher and Jonathan Mak for stimulating discussions. This work was funded by the Deutsche Forschungsgemeinschaft (DFG, German Research Foundation) – Project-ID 328545488 – TRR 227, project B03.

\begin{appendix}

\section{Simple model for easy-plane anisotropy}
\label{sec:anisotropy}
This Appendix gives an explicit expression for the response time $\tau$ in the case that the anisotropy energy of the magnet is quadratic in $M_z$. 
The helical edge state is described by the second-quantized Hamiltonian \cite{Meng2014-rs, Silvestrov2016-sp}
\begin{equation}
    \mathcal{H} = \int \mathrm{d}x \, \hat{\boldsymbol{\psi}}_x^\dagger \left[ - i \hbar v_{\rm F} \partial_x \sigma_z + h(x) \boldsymbol{\sigma} \cdot \mathbf{M} \right] \hat{\boldsymbol{\psi}}_x + \frac{1}{2}\frac{D}{V_{\rm M}} M_z^2,
\label{eq:hamiltonian}
\end{equation}
where $\hat{\boldsymbol{\psi}}_x = (\hat{\psi}_\uparrow(x), \hat{\psi}_\downarrow(x))^\mathrm{T}$ is a two-component spinor of the helical edge states, $v_{\rm F}$ the Fermi velocity of the electrons, $h(x)$ a spatial profile that determines the interaction with the magnet, and the last term denotes the easy-plane anisotropy with $D > 0$. Furthermore we introduced the gyromagnetic ratio $\gamma < 0$ and the volume of the magnet $V_{\rm M}$. In our notation, the symbol $\vM$ denotes the magnetic moment of the magnet. In Ref.\ \onlinecite{Silvestrov2016-sp}, this symbol is used to denote the ``angular momentum'' (measured in units of $\hbar$), which is equal to $\vM/\gamma \hbar$ in our notation. Correspondingly the spatial profile function $h(x)$ and the easy-plane anisotropy constant $D$ used in Ref.\ \onlinecite{Silvestrov2016-sp} are equal to $h(x) \hbar\gamma$ and $D (\hbar \gamma)^2/V_{\rm M}$ in our notation, respectively. In Ref.\ \onlinecite{Meng2014-rs}, the symbol $\vM$ is used to denote the magnetization, which is equal to $\vM/V_{\rm M}$ in the units used here.

We describe the magnet using the macrospin approximation, in which the dynamics of the magnetic moment follow the Landau–Lifshitz-Gilbert equation \cite{Foros2005-px}
\begin{equation}
    \dot{\mathbf{M}} = \gamma \mathbf{M} \times \left( - \frac{\partial \mathcal{H}}{\partial \mathbf{M}} \right) + \gamma \mathbf{M} \times \mathbf{h}(t) + \frac{\alpha}{M_{\rm s}} \mathbf{M} \times \dot{\mathbf{M}}.
\label{eq:ll}
\end{equation}
For the contribution from the derivative of the Hamiltonian in Eq.~\eqref{eq:ll}, we treat the contribution from the interaction of the magnet with the conduction electrons in the scattering framework, see Eq.\ (\ref{eq:dotMz}) of the main text.
The stochastic-field term (with $\langle \mathbf{h}(t) \rangle = 0$) and the Gilbert damping term (with Gilbert damping parameter $\alpha$) are disregarded in the main text, but will be considered in App.~\ref{sec:heatbath}.

A harmonic {\it Ansatz} for the precession of the magnetic moment around the $z$ axis with frequency $\partial_t \varphi(t) = \Omega(t)$ and small out-of-plane canting angle $|\theta(t)| \ll 1$, {\it i.e.}, $M_x = M_{\rm s} \cos{\varphi (t)}$, $M_y = M_{\rm s} \sin{\varphi (t)}$ and $M_z = M_{\rm s} \sin{\theta(t)} $, solves Eq.~\eqref{eq:ll} \cite{Meng2014-rs}. The {\it Ansatz} implies for the $z$ component of the magnetic moment
\begin{equation}
    M_z = \frac{V_{\rm M}}{\gamma D} \Omega ,
\end{equation}
which establishes a relation between precession frequency $\Omega$ and canting angle $\theta$,
\begin{equation}
    \Omega(t) = \frac{\gamma D M_{\rm s}}{V_{\rm M}} \sin{\theta(t)}.
\end{equation}
We find that for the anisotropy specified in Eq.\ \eqref{eq:hamiltonian} the response time given in Eq.\ \eqref{eq:tau} simplifies to ({\it cf.} Ref.~\cite{Meng2014-rs})
\begin{equation}
    \tau = \frac{2 \pi V_{\rm M}}{\hbar \gamma^2 D \mathcal{R}}.
\end{equation}
The case of larger out-of-plane canting angles, which is relevant in the presence of a DC voltage bias, is discussed in App.\ \ref{sec:dcbias}.

\section{Coupling of magnet to a heat bath}
\label{sec:heatbath}
In the main text, we only consider the coupling to electrons of the helical edge state as a source of relaxation and of thermal fluctuations of the magnetic moment. In this appendix, we also take into account an additional coupling of the magnet to a thermostat, {\it e.g.}, phonons at a finite temperature.

To this end, following Ref.~\onlinecite{Meng2014-rs}, we add two terms to Eq.~\eqref{eq:dotMzz} of the main text: a damping term proportional to a dimensionless Gilbert damping parameter $\alpha$ and a stochastic term $h(t)$, which is related to the damping term via the fluctuation-dissipation theorem \cite{Callen1951-ho, Foros2005-px},
\begin{align}
    \langle h(\omega) \rangle &= 0, \nonumber \\
    \langle h(\omega) h(\omega') \rangle &= \frac{4 \pi \hbar \omega \alpha }{\gamma M_{\rm s}} \delta(\omega + \omega') \coth{\frac{\hbar \omega}{2 k_{\rm B} T}}.
\end{align}
Inclusion of these additional contributions into Eq.~\eqref{eq:dotMzz} gives
\begin{align}
  \dot{M_z}(t) =&\
  \frac{\gamma}{4 \pi}
  \int \mathrm{d}\varepsilon \int \mathrm{d}\varepsilon' 
  e^{i (\varepsilon - \varepsilon')t/\hbar} \nonumber \\
  \begin{split}
  &\, \times 
  [\hat a^{\dagger}_{\rm L}(\varepsilon) \hat a_{\rm L}(\varepsilon')
  + \hat b^{\dagger}_{\rm L}(\varepsilon) \hat b_{\rm L}(\varepsilon') \\ 
  &\, \quad \ - \hat a^{\dagger}_{\rm R}(\varepsilon) \hat a_{\rm R}(\varepsilon')
  - \hat b^{\dagger}_{\rm R}(\varepsilon) \hat b_{\rm R}(\varepsilon')]
   \end{split}  \label{eq:dotMzz_heatbath} \\
  &\, - \frac{\hbar \gamma}{e} \hat I_{\rm pump}(t)
  \nonumber \\ &\, \mbox{}
  + \gamma M_{\rm s} h(t) - \alpha M_z(t) \left( \frac{\mathrm{d} \Omega}{\mathrm{d} \theta} \right)_{\!\! \Omega = 0} . \nonumber
\end{align}
Solving the coupled equations (\ref{eq:Ipump}), (\ref{eq:dotMzz_heatbath}), and (\ref{eq:MzOmega}) in frequency space gives an expression for the pumped current $\hat I_{\rm pump}$ in the presence of this additional source of relaxation,
\begin{align}
\label{eq:IpumpResult_heatbath}
  \hat I_{\rm pump}(\omega)=&\ \frac{e}{2(1 - i \omega \tilde \tau)} \frac{\tilde \tau}{\tau}
  \bigg[ \frac{2 M_{\rm s}}{\hbar} h(\omega) \\
  &\, + \int \mathrm{d}\varepsilon
  [ \hat a^{\dagger}_{\rm L}(\varepsilon) \hat a_{\rm L}(\varepsilon+\hbar \omega)
  + \hat b^{\dagger}_{\rm L}(\varepsilon) \hat b_{\rm L}(\varepsilon+\hbar \omega) \nonumber \\ &\, \quad \
  - \hat a^{\dagger}_{\rm R}(\varepsilon) \hat a_{\rm R}(\varepsilon+\hbar \omega)
  - \hat b^{\dagger}_{\rm R}(\varepsilon) \hat b_{\rm R}(\varepsilon+\hbar \omega) ] \bigg], \nonumber
\end{align}
where $\tau$ was defined in Eq.~\eqref{eq:tau} and 
\begin{align}
    \frac{1}{\tilde \tau} =&\, \frac{1}{\tau} + \frac{1}{\tau_{\alpha}}
\label{eq:tau_tilde}
\end{align}
with an effective relaxation rate
\begin{align}
  \frac{1}{\tau_{\alpha}} = 
  \alpha \left( \frac{\mathrm{d} \Omega}{\mathrm{d} \theta} \right)_{\!\! \theta \to 0}
\end{align}
that combines contributions from spin pumping and Gilbert damping \cite{Tserkovnyak2002-ax,Meng2014-rs}.
For the equilibrium noise, see Eq.~\eqref{eq:noisepower}, we find
\begin{align}
\label{eq:S_heatbath}
  S_0(\omega) =&\, \frac{2 e^2}{h} \left( \hbar \omega  \coth{\frac{\hbar \omega}{2 k_{\rm B} T}} \right)
  \bigg\{ \frac{\tilde \tau^2}{\tau^2} \frac{1}{1 + \omega^2 \tilde \tau^2} \frac{2 \pi \alpha M_{\rm s}}{\hbar \gamma} \nonumber \\
  &\, \mbox{} + \int \mathrm{d}\varepsilon \left[|t(\varepsilon)|^2 + \frac{\tilde \tau^2}{\tau^2} \frac{1}{1 + \omega^2 \tilde \tau^2} |r(\varepsilon)|^2 \right] \nonumber \\
  &\, \ \ \ \ \times \frac{f(\varepsilon) - f(\varepsilon + \hbar \omega)}{\hbar \omega} \bigg\} .
\end{align}
We may again simplify this result making use of the limit $|\omega| \ll \omega_{\rm Th}$, where we find
\begin{align}
\label{eq:S_heatbath_simplified}
    S_0(\omega) =&\ \frac{e^2 \omega}{\pi}
  \left( 1 - \mathcal{R} \frac{\tfrac{\tau - \tilde \tau}{\tau} + \omega^2 \tilde \tau^2}{1 + \omega^2 \tilde \tau^2} \right) \coth{\frac{\hbar \omega}{2 k_{\rm B} T}} .
\end{align}
One verifies that Eq.~(\ref{eq:S_heatbath_simplified}) agrees with Eq.~\eqref{eq:Sfinal} in the limit $\alpha \to 0$. 

For comparison, we also calculate the AC conductance in the presence of Gilbert damping. From Eqs.~\eqref{eq:IL}, \eqref{eq:aaomega} and \eqref{eq:IpumpResult_heatbath} we find 
\begin{align}
  \label{eq:Gomegaalpha}
    G_0(\omega) = &\,
    \frac{e^2}{h} \int \mathrm{d}\varepsilon \frac{f(\varepsilon)-f(\varepsilon+\hbar \omega)}{\hbar \omega} \nonumber \\ &\, \mbox{} \times
    \left[ |t(\varepsilon)|^2 + \frac{\tilde \tau}{\tau} \frac{1}{1-i\omega\tilde \tau} |r(\varepsilon)|^2\right] \nonumber \\ =&\, 
    \frac{e^2}{h} \left(1 - \mathcal{R} \frac{\tfrac{\tau - \tilde \tau}{\tau} - i \omega \tilde \tau}{1 - i \omega \tilde \tau} \right),
\end{align}
where, in the second equality, we again made use of the limit $|\omega| \ll \omega_{\rm Th}$. One easily verifies that Eqs.\ (\ref{eq:S_heatbath_simplified}) and (\ref{eq:Gomegaalpha}) again satisfy the fluctuation-dissipation relation \eqref{eq:fdt}.

For a macroscopic magnet with nonzero Gilbert damping, one typically has $\tilde \tau \ll \tau$, see Eq.~(\ref{eq:tau_tilde}) \cite{Meng2014-rs}. The Gilbert damping prevents the current-driven precession of the magnetic moment, so that there is no pumped spin current and the conductance equals that of an elastic scatterer, $G_0(\omega) = (e^2/h)(1-\mathcal{R})$, independent of frequency. The noise power is affected correspondingly.
With additional Gilbert damping, the precession frequency for finite DC bias $V$ is no longer given by $\Omega_V = e V/\hbar$. As a consequence, the arguments given in App.\ \ref{sec:dcbias} do not apply and one can not conclude that the noise power is bias-independent.

\section{Bias dependence of the noise}
\label{sec:dcbias}

To obtain Eq.~(\ref{eq:SV}) of the main text, we recall that the DC bias $V$ causes the magnetic moment to precess at frequency $\Omega_V = eV/\hbar$ and with out-of-plane canting angle $\theta_V$. Current fluctuations impose small deviations of the precession frequency $\Omega$ and the canting angle $\theta$ from their steady-state values $\Omega_V$ and $\theta_V$. To zoom in on these fluctuations, we transform to a spin reference frame that rotates around $\ve_z$ with frequency $\Omega_V$. In the rotating frame, the magnetic moment precesses with frequency $\Omega' = \Omega - \Omega_V$ around $\ve_z$. The precession frequency $\Omega'$ in the rotating frame has small fluctuations around $\Omega' = 0$. To describe these fluctuations, we repeat the calculations of Sec.\ \ref{sec:magnethelicaledge} for the rotating frame. 

Creation and annihilation operators in the rotating frame and in the original frame are related as
\begin{align}
\begin{split}
  \hat a_{\rm L}'(\varepsilon) =&\, \hat a_{\rm L}(\varepsilon + \tfrac12 eV), \\
  \hat a_{\rm R}'(\varepsilon) =&\, \hat a_{\rm R}(\varepsilon - \tfrac12 eV), \\
  \hat b_{\rm L}'^{\Omega'}(\varepsilon) =&\, \hat b_{\rm L}^{\Omega}(\varepsilon - \tfrac12 eV), \\
  \hat b_{\rm R}'^{\Omega'}(\varepsilon) =&\, \hat b_{\rm R}^{\Omega}(\varepsilon + \tfrac12 eV),
\end{split}
\label{eq:abtransform}
\end{align}
where we use a prime to denote operators in the rotating frame. As in Sec.\ \ref{sec:magnethelicaledge}, the superscripts $\Omega'$, $\Omega$ for the operators $\hat b'_{\rm L,R}$, $\hat b_{\rm L,R}$ for outgoing scattering states indicate that these are defined for a magnetic moment precessing at frequency $\Omega'$ or $\Omega$, respectively.
The transformation rules (\ref{eq:abtransform}) imply that the applied bias voltage in the rotating frame $V' = 0$. The relation between operators for outgoing and incoming scattering states in the rotating frame reads
\begin{align}
  \label{eq:abrotating}
\begin{split}
  \hat b_{\rm L}'^{\Omega'}(\varepsilon-\tfrac12 \hbar \Omega') &=
  r(\varepsilon) \hat a_{\rm L}'(\varepsilon + \tfrac12 \hbar \Omega')
  + t'(\varepsilon) \hat a_{\rm R}'(\varepsilon-\tfrac12 \hbar \Omega'), \\
  \hat b_{\rm R}'^{\Omega'}(\varepsilon+\tfrac12 \hbar \Omega') &=
  t(\varepsilon) \hat a_{\rm L}'(\varepsilon+\tfrac12 \hbar \Omega')
  + r'(\varepsilon) \hat a_{\rm R}'(\varepsilon - \tfrac12 \hbar \Omega').
\end{split}
\end{align}
The energy shifts $\pm (1/2) \hbar \Omega'$ appearing in Eq.\ (\ref{eq:abrotating}) are of the order of the fluctuations of the precession frequency $\Omega$ and no longer of the order of $\Omega$ itself. This means that the approximations made in Sec.\ \ref{sec:magnethelicaledge} can also be made in the rotating frame, because any precession-induced shifts of the energy arguments of the reflection and transmission amplitudes are small in comparison to the Thouless frequency $\omega_{\rm Th}$.

The current operators $\hat I'_{\rm L,R}$ in the rotating frame are 
\begin{align}
  \label{eq:Itransform}
\begin{split}
  \hat I'_{\rm L,R}(t) =&\ 
   \frac{e}{h} \int \mathrm{d}\varepsilon \int \mathrm{d}\varepsilon'
  e^{i(\varepsilon - \varepsilon') t / \hbar} \\ 
  &\, \times
  [\hat a_{\rm L,R}^{\dagger}(\varepsilon) \hat a_{\rm L,R}(\varepsilon')
  - \hat b_{\rm L,R}'^{\dagger}(\varepsilon) 
  \hat b'_{\rm L,R}(\varepsilon')] \\
  &\, \mbox{} + \hat I_{\rm pump}'(t) + \frac{e^2 V}{h}.
\end{split}
\end{align}
Here we expanded to linear order in the small frequency $\Omega'$, similar as in Eq.~(\ref{eq:IL}), wrote $\hat b_{\rm L,R}'(\varepsilon) = \hat b_{\rm L,R}'^{\Omega' = 0}(\varepsilon)$, and introduced
\begin{equation}
  \hat I_{\rm pump}'(t) = \frac{e}{2 \pi} \Omega' \mathcal{R}.
\end{equation}
The reflection coefficient $\mathcal{R}$ is calculated in the rotating frame using Eq.\ (\ref{eq:r}). In the original frame, one may also use Eq.\ (\ref{eq:r}) to calculate ${\cal R}$ if the energy arguments of the reflection and transmission amplitudes are defined according to Eq.\ (\ref{eq:abrt}).
Apart from the constant offset $e^2 V/h$, Eq.~(\ref{eq:Itransform}) is formally identical to the expression (\ref{eq:IL}) for the fluctuating current in equilibrium.

The transformation to the rotating frame does not affect the out-of-plane canting angle $\theta$ and the out-of-plane component $M_z$ of the magnetization. To account for small fluctuations of $\theta$ and $M_z$ around their steady-state value $\theta_V$ and $M_{\rm s} \sin \theta_V$, we write
\begin{equation}
  M_z = M_{\rm s} \sin \theta_V + M_z',
\end{equation}
so that
\begin{equation}
  \Omega' = \frac{M_z'}{M_{\rm s} \cos \theta_V}
  \left( \frac{d\Omega}{d \theta} \right)_{\!\! \theta = \theta_V}.
  \label{eq:Omegaprime}
\end{equation}
Calculating $\dot M_z'$ in the rotating frame and expanding in small $\Omega'$ gives (compare with Eqs.\ (\ref{eq:dotMz}) and (\ref{eq:dotMzz}))
\begin{align}
\begin{split}
  \dot{M_z}(t) =&\
  \frac{\gamma}{4 \pi}
  \int \mathrm{d}\varepsilon \int \mathrm{d}\varepsilon' 
  e^{i (\varepsilon - \varepsilon')t/\hbar} \\ &\, \times 
  [\hat a'^{\dagger}_{\rm L}(\varepsilon) \hat a'_{\rm L}(\varepsilon')
  + \hat b'^{\dagger}_{\rm L}(\varepsilon) \hat b'_{\rm L}(\varepsilon') \\
  &\, \quad \ - \hat a'^{\dagger}_{\rm R}(\varepsilon) \hat a'_{\rm R}(\varepsilon')
  - \hat b'^{\dagger}_{\rm R}(\varepsilon) \hat b'_{\rm R}(\varepsilon')]
  \nonumber \\ &\, \mbox{}
  - \frac{\hbar \gamma}{e} \hat I_{\rm pump}'(t).
\end{split}
\label{eq:dotMzV}
\end{align}

Apart from the offset $e^2 V/h$ in Eq.~(\ref{eq:Itransform}) and a different proportionality constant in the relation (\ref{eq:Omegaprime}) between $\Omega'$ and $M_z'$, these equations are precisely the same as the equations (\ref{eq:IL}), (\ref{eq:Ipump}), (\ref{eq:dotMzz}), and (\ref{eq:MzOmega}) that govern the current and the current fluctuations in equilibrium. We may then repeat the calculations of Secs.\ \ref{sec:magnethelicaledge}--\ref{sec:noise} to find the admittance $G_V(\omega)$ and noise power $S_V(\omega)$. The results are the same as those for $G_0(\omega)$ and noise power $S_0(\omega)$, except for the replacement $\tau \to \tau_V$.

We note that the above derivation does not require that the actual precession frequency $\Omega$ to be small in comparison to the Thouless frequency $\omega_{\rm Th}$; it only requires that the difference $\Omega' = \Omega - \Omega_V$ be $\ll \omega_{\rm Th}$. Smallness of $\Omega'$ can be verified in the same way as was done in Sec.\ \ref{sec:noise} for the smallness of the precession frequency in the absence of an applied bias.

\end{appendix}

\bibliographystyle{apsrev4-1}
\bibliography{main}

\end{document}